\documentclass[pre,amsmath,amssymb,aps,twocolumn,superscriptaddress,longbibliography]{revtex4-2}
\usepackage{amssymb}
\usepackage{amsmath}
\usepackage{commath}
\usepackage{verbatim}
\usepackage{graphicx}
\usepackage{dcolumn}
\usepackage{bm}
\usepackage[dvipsnames]{xcolor}
\usepackage[colorlinks]{hyperref}
\usepackage{graphicx}
\usepackage{array}
\usepackage{physics}
\newcolumntype{o}{@{}>{{}}c<{{}}@{}}
\hypersetup{                
	colorlinks=true,                
	breaklinks=true,                
	urlcolor= blue,                
	linkcolor= red,                
    bookmarksopen=false,
	filecolor=black,
	citecolor=blue,
	linkbordercolor=black
}

\graphicspath{ {images/} }
\begin{document}

\title{Abrupt transitions in the optimization of diffusion with distributed resetting}

\author{Pedro Juli\'an-Salgado}
 \email{pjuliansalgado@gmail.com}
\affiliation{Physics Department, Universidad Autónoma Metropolitana, Mexico City 09340, Mexico}%
\author{Leonardo Dagdug}%
 \email{dll@xanum.uam.mx}
\affiliation{Physics Department, Universidad Autónoma Metropolitana, Mexico City 09340, Mexico}%
\author{Denis Boyer}
 \email{boyer@fisica.unam.mx}
\affiliation{Instituto de Física, Universidad Nacional Autónoma de México, Mexico City 04510, Mexico}


\begin{abstract}
Brownian diffusion subject to stochastic resetting to a fixed position has been widely studied for applications to random search processes. In an unbounded domain, the mean first-passage time at a target site can be minimized for a convenient choice of the resetting rate. Here we study this optimization problem in one dimension when resetting occurs to random positions, chosen from a probability density function with compact support that does not include the target. Depending on the shape of this distribution, the optimal resetting rate either varies smoothly with the mean distance to the target, as in single-site resetting, or exhibits a discontinuity caused by the presence of a second local minimum in the mean first-passage time. 
These two regimes are separated by a critical line containing a singular point that we characterize through a Ginzburg-Landau theory. To quantify how useful is a given resetting point for the search, we calculate the probability density function of the last resetting position before absorption. The discontinuous transition separates two markedly different optimal strategies: one with a small resetting rate where the last path before absorption starts from a rather distant but likely position, while the other strategy has a large resetting rate, favoring last paths starting from not-so-likely points but which are closer to the target.

\end{abstract}

\maketitle
\section{Introduction}
Diffusive processes subject to stochastic resetting to a given position represent an efficient way of carrying out a random search task. After some time of unfruitful search to locate a specific target region, resetting gives the opportunity to the searcher to come back to its starting point for exploring new pathways. Brownian motion under resetting achieves its first passage at a target site in an unbounded medium within a finite time, contrary to ordinary diffusion where such a time diverges on average. In this situation, which is quite generic, the mean first-passage time (MFPT) is a non-monotonous function of the resetting rate, with a unique minimum at a certain optimal value \cite{evans_diffusion_2011,evans_diffusion_2011-1,evans2014diffusion,pal_diffusion_2016,chechkin2018random,evans_stochastic_2020}. This property makes resetting appealing as a mechanism to expedite the completion of a task, and applications have been discussed in a variety of biological contexts, for instance, enzymatic reaction kinetics \cite{reuveni2014role,rotbart2015michaelis, pal2019landau}, RNA recovery \cite{roldan2016stochastic}, active cellular transport \cite{bressloff_modeling_2020}, animal foraging \cite{mercado2018lotka,giuggioli2019comparison,pal2020search,altshuler2024environmental} or  population dynamics \cite{kang2022evolutionary,Asymmetric_Plata_Carlos,ali_asymmetric_2022}. The optimization of first passage times through resetting has also been generalized to a variety of other processes such as anomalous diffusion models \cite{kusmierz2014first,kusmierz2015optimal, kusmierz2019subdiffusive,mendez2021continuous,pal2024random}, and tested on Brownian motion in controlled experiments using colloidal particles \cite{besga_optimal_2020,faisant_optimal_2021,friedman_resetting_experiment}.

Nevertheless, the effect of resetting on first passage times can be diverse. Resetting can also hinder the completion of a task instead of expediting it, as in some cases the MFPT turns out to be a monotonously increasing function of the resetting rate $r$. A simple universal criterion actually predicts the effect of a small resetting rate on the MFPT of an arbitrary process \cite{reuveni_optimal_2016,pal2017first}. An example is provided by a Brownian particle in an interval: depending on the resetting/starting position, the mean exit time from the interval is either a non-monotonous function of $r$ with a minimum at $r^{*}>0$, or a monotonously increasing function of $r$ whose optimum is therefore $r^*=0$ (the absence of resetting favors the exit) \cite{pal2019first,durang2019first, singh2020resetting}. The transition between these two optimal regimes occurs at a certain resetting position and it is of \lq\lq second order", i.e., $r^*$ tends to $0$ continuously. A similar transition occurs in other setups, e.g., in drift-diffusion processes with an absorbing boundary \cite{ray2019peclet}.

In some systems, the variations of the MFPT with the resetting rate $r$ are more complex and exhibit two local minima, one at a value $r_1>0$ and the other at $r_2=0$. Examples range from the kinetics of enzymatic catalysis \cite{rotbart2015michaelis,pal2019landau} to Brownian motion under resetting in certain external potentials \cite{ahmad2022first} or diffusion in narrow channels of varying width \cite{jain2023fick}. In these cases, discontinuous (or \lq\lq first order") transitions from a vanishing to a non-zero optimal resetting rate can be observed as the resetting position (or some other parameter) is varied. Another example where the MFPT at a target site can exhibit more than one minimum is provided by Brownian diffusion in unbounded space with resetting to a random position taken from a Gaussian distribution \cite{besga_optimal_2020,faisant_optimal_2021}. Since the particle has a non-zero probability density to be reset at the target position itself, the MFPT is usually a decreasing function of $r$ in that case, and the globally optimal rate is always $r_1=\infty$. However, if the variance of the Gaussian is sufficiently small, a secondary \lq\lq metastable" minimum at a finite $r_2$ appears.  

A situation of particular interest in the present study and which has been little discussed previously, is the case of a MFPT having multiple minima at some finite rates, e.g., $0<r_2<r_1<\infty$, meaning that both a fast and a slow resetting strategy are locally optimal, while neither $r=0$ nor $r=\infty$ are advantageous. The mean search time of symmetric L\'evy flights of index $\mu$ under resetting exhibits such a behavior, with two local minima in the $(r,\mu)$ plane \cite{kusmierz2015optimal}. A more recent example is Brownian search under a time constraint and where each resetting event comes with a cost, such that the searcher seeks to minimize the total cost while maintaining a high probability of locating the target within the allotted time \cite{sunil2024minimizing}.  
In the context of quantum walks, mean detection times with multiple minima have been found in evolutions subject to periodic resetting \cite{yin2023restart,yin2024instability}. A fourth example is diffusion with two resetting points \cite{two_resetting_points}, a simple extension of the paradigmatic single-site resetting search problem of
\cite{evans_diffusion_2011}. All these examples exhibit rich behaviors for the optimal strategy $r^*$, in particular discontinuous transitions between finite values.

In this paper, we revisit the first passage properties of diffusive processes under resetting to randomly distributed positions and show that they naturally exhibit discontinuous transitions when the resetting positions have a probability density function (PDF) with compact support, thus generalizing our previous finding on two resetting points \cite{two_resetting_points}. Resetting to multiple sites finds applications in many fields such as foraging ecology \cite{lambertucci2010size,lambertucci2013cliffs}, human mobility \cite{song2010limits}, internet search \cite{brin1998anatomy,ermann2015google}, intermittent target searches along the DNA \cite{coppey2004kinetics}, and can also be implemented in optical tweezers experiments \cite{besga_optimal_2020,faisant_optimal_2021}. A handful of works have examined the effects of stochastic resetting to a distribution of positions, since the initial work of Evans and Majumdar \cite{evans_diffusion_2011-1}. For example \cite{gonzalez_diffusive_2021} investigated the impact of resetting to multiple nodes on first passage times for random walks on networks.  Depending on the number of resetting nodes, the mean first passage time may or may not be optimizable by a finite resetting probability.
Ref. \cite{olsen_steady-state_2023} developed a method for calculating the moments of the steady-state distribution of a process with resetting, expressed in terms of averages computed from the resetting distribution. 
Distributed resetting was also studied from the point of view of information theory \cite{toledo2023first}.
In \cite{mendez_first_passage}, the mean time to absorption was analyzed under different resetting scenarios where the optimal resetting rate varied smoothly. 

Most of the above theoretical studies have considered a resetting point PDF with infinite support, such that it is always possible for the searcher to reset arbitrarily close to the target (although with a small probability). However, in practical applications, the target may not be part of the possible resetting positions, which often are distributed on a finite support. For instance, optical tweezers create a trapping zone of finite range \cite{pesce2020optical}; animals like condors may use several specific roosts as resetting sites as they forage in a much wider environment \cite{lambertucci2010size,lambertucci2013cliffs}. Likewise, models of human mobility consider a few places (such as home and workplace) as fixed positions around which less predictable displacements occur \cite{song2010limits,gonzalez_diffusive_2021}.

The paper is organized as follows. We first recall the theory of distributed resetting and show the theoretical interest of considering truncated PDFs, i.e., restricted to a compact support (Section \ref{sec:setup}). We apply these results to truncated Gaussian and exponential distributions, and unveil abrupt transitions in the optimal rate as the mean or variance are varied (Section \ref{sec:other_rho}). In Section \ref{sec:GL}, we develop a Ginzburg-Landau theory that accounts phenomenologically for these first-order transitions and derive a general condition for the existence of an associated critical point, in analogy with the liquid-gas critical point. This singularity with infinite derivative is located on a critical line where the variations of $r^*$ are still continuous. To obtain further insights on the mechanisms that lead to a discontinuous transition, we study in Section \ref{sec:lastresetting} the {\em last resetting point probability density function}, i.e., the PDF of the last resetting position used by the particle before finding the target.  
The analysis of this distribution allows us to show how two equivalent, \lq\lq coexisting" optimal strategies can be markedly different. We also revisit the two-resetting point problem under this scope and see how it differs in some aspects from the other distributions studied. We conclude in Section \ref{sec:concl}.

\section{Setup}\label{sec:setup}
Let us consider Brownian diffusion subject to stochastic resetting with rate $r$ to random positions $z$ distributed with a PDF $\rho(z)$ \cite{evans_diffusion_2011-1}. More specifically, during the time interval $[t,t+\dd t]$, with probability $r\dd t$, a position $z$ is drawn from the distribution $\rho(z)$ and the particle is set to position $z$, while with complementary probability $1-r\dd t$, the particle continues to diffuse. The successive resetting positions are assumed to be independent.

\begin{figure}
\centering
\includegraphics[width=\linewidth]{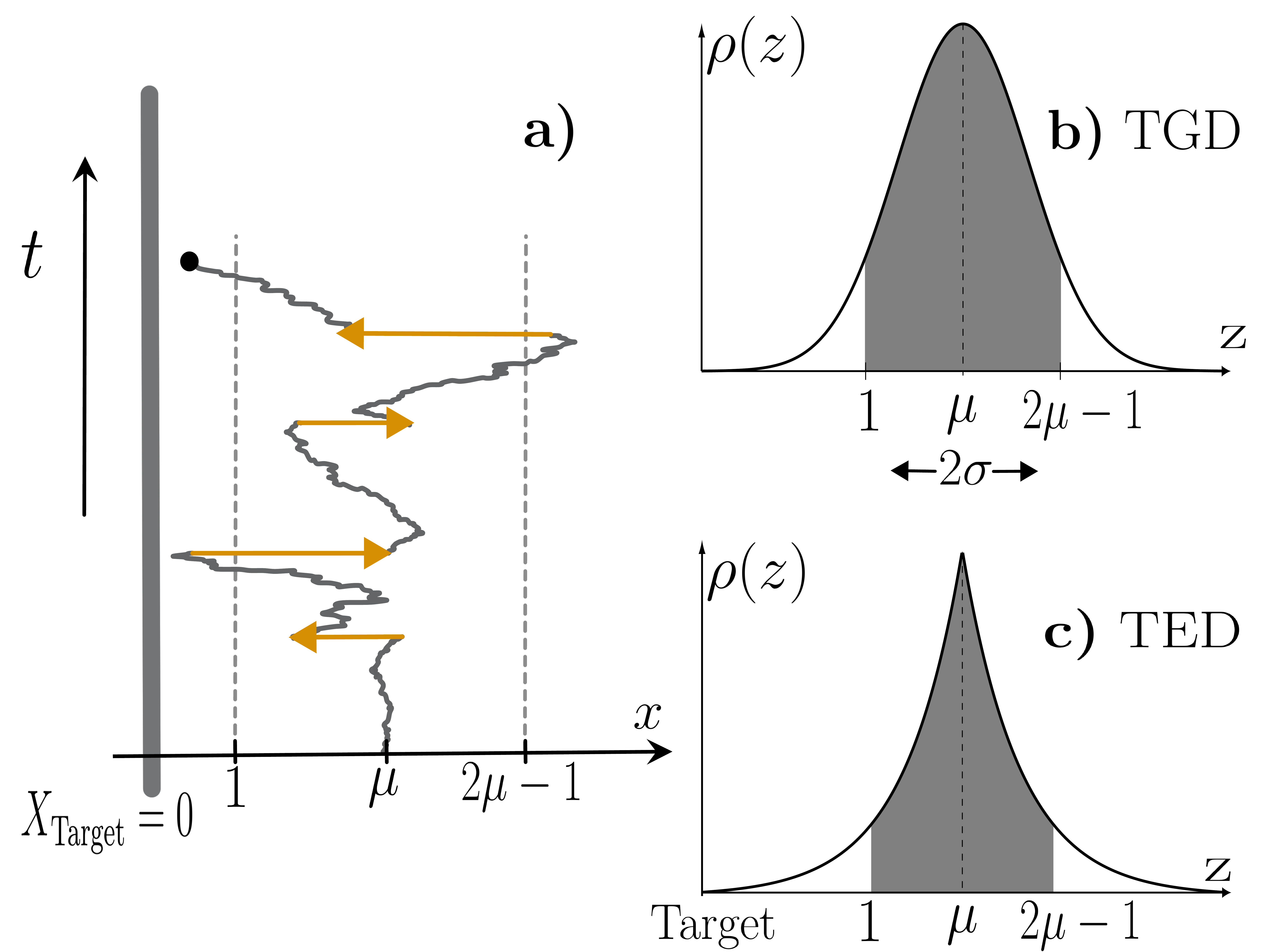}
  \caption{(a) Sample path of a $1d$ diffusive particle subject to resetting at rate $r$ to random positions $z$, with an absorbing wall located at $x=0$.  At each reset, the particle is set to an independent position $z$ distributed with the PDF $\rho(z)$. In gray filling, two examples of distribution for $z$: (b) truncated Gaussian distribution (TGD), and (c) truncated exponential distribution (TED), both with mean $\mu$. In these cases, the support of $\rho(z)$ is $z\in[1,2\mu-1]$, while $\rho(z)=0$ outside.}\label{fig:illustration}
\end{figure}

Here, the motion takes place in a one-dimensional, semi-infinite channel with a single absorbing target at the origin ($x = 0$), see Fig. \ref{fig:illustration}a.
We recall the general renewal equation for the survival probability \cite{evans_diffusion_2011-1},
\begin{multline}
S_{r}(t|x_{0})=e^{-rt}S_{0}(t|x_{0})\\+\int_{-\infty}^{\infty} \rho (z)\left[ r\int_{0}^{t}\dd%
\tau e^{-r\tau }S_{r}(t-\tau |x_{0})S_{0}(\tau |z)\right] \dd z,
\label{pdf-1}
\end{multline}
where $S_{r}(t|x_{0})$ stands for the probability that the particle subject to resetting and with initial position $x_0$ at $t=0$ has not crossed the origin until time $t$, while $S_{0}(t|x_{0})$ is the survival probability in the absence of resetting ($r=0$). For the free Brownian particle, $S_{0}(t|x_{0})={\rm erf}(\frac{|x_{0}|}{\sqrt{4Dt}})$ and $D$ is the diffusion coefficient. 

The first term of the right-hand side (rhs) of Eq. \eqref{pdf-1} comes from the surviving trajectories that have not been reset until time $t$, which occurs with probability $e^{-rt}$. The second term in the rhs is an average over the resetting position $z$ where the particle had its last reset before the current time $t$. This average applies to trajectories that had their last reset at some time $t-\tau$ and then continued to evolve during the interval $[t-\tau,t]$ with a simple Brownian motion. 
We proceed further by applying the Laplace transform ${\cal L}(\cdot)=\int_0^{\infty}e^{-st}(\cdot)dt$ to both sides of Eq. \eqref{pdf-1}, which reduces the problem to an algebraic equation whose solution is
\begin{equation}
q_{r}(s|x_{0})=\frac{q_{0}(r+s|x_{0})}{1-r\int_{-\infty}^{\infty} \rho (z)q_{0}(r+s|z)\dd z},  \label{pdf-2}
\end{equation}
where $q_{r,0}(s|x_i)={\cal L}[S_{r,0}(t|x_i)]$, respectively. For free Brownian motion, one can rely on a specific closed form given by $q_{0}(s|x_i) = \frac{1-e^{-\sqrt{s/D}|x_i|}}{s}$, see, e.g. \cite{evans_diffusion_2011}. From Eq. \eqref{pdf-2} the MFPT for a particle starting from $x_0$ is readily identified with $q_{r}(s=0|x_{0})=\int_0^{\infty} dt\ S_r(t,x_0)$ \cite{dagdug_diffusion_2024} and given by 
\begin{equation}
T(x_{0}) =\frac{1-e^{-\alpha_{0} |x_{0}|}}{r\int_{-\infty}^{\infty} \rho \left(z\right) e^{-\alpha_{0} |z|}\dd z} \label{pdf-3},
\end{equation}
where 
\begin{equation}\label{alpha0}
\alpha_0 = \sqrt{r/D}
\end{equation}
is proportional to the inverse typical distance diffused by the particle between two consecutive resets.
    
In the following, we assume that the initial position $x_0$ is also distributed according to $\rho(z)$. The main quantity of interest is therefore the averaged mean first-passage time (AMFPT), defined by 

\begin{equation}\label{amfpt}    
    \mathcal{T}_{r}=\int\rho(x_0)T(x_0)\dd x_0 
\end{equation}  

and whose expression is
\begin{equation}
	\mathcal{T}_{r}=\frac{1}{r}\left( \frac{1}{\int_{-\infty}^{\infty} \rho \left(
		z\right) e^{-\alpha_{0} |z|}\dd z}-1\right). \label{pdf-4}  
\end{equation}

The purpose of the present study is to determine the rate $r^*$ where Eq. (\ref{pdf-4}) reaches a global minimum. 
If $\rho(z)=\delta(z-\mu)$ with $\mu$ a constant, one recovers the well-known single-site resetting problem, for which ${\cal T}_r$ has a unique minimum, at $r^*\simeq 2.5396\ D/\mu^2$ \cite{evans_diffusion_2011}.
More generally, at small $r$, it is easy to see that Eq. (\ref{pdf-4}) yields 
\begin{equation}
{\cal T}_r\simeq \frac{\langle|z|\rangle}{\sqrt{Dr}}\to\infty,
\label{smallr}
\end{equation}
    similarly to single-site resetting, where the average $\langle|z|\rangle$ with weight $\rho(z)$ is replaced by $\mu$ \cite{evans_diffusion_2011}. 

    Let us now point out two rather different situations related to the {\em large} $r$ behavior of ${\cal T}_r$:
    
    \hspace{0.5cm}$\bullet$ Case $\rho(0)\neq 0$, i.e., the  PDF at the target position is finite. In the large $r$ limit (taking $\alpha_0\to\infty$), $\rho(z)$ can be replaced by $\rho(0)$ in Eq. (\ref{pdf-4}) and integration over $z$ gives
\begin{equation}
\mathcal{T}_{r}\simeq \frac{1}{2\rho(0)\sqrt{Dr}}\to0\, .
\label{Trcase1}
\end{equation}
    Since $\lim_{r\to\infty}{\cal T}_r=0$, the optimal resetting rate $r^*$ is infinite. Qualitatively, when the distribution $\rho(z)$ is sampled infinitely fast, the particle can reset very near the target site and the latter is quickly found after very little diffusion. Although ${\cal T}_r$ behaves as $1/\sqrt{r}$ for both large and small values of $r$, it can exhibit non-monotonous variations for intermediate values of $r$, depending on the variance of $\rho(z)$. This scenario was studied in \cite{besga_optimal_2020,faisant_optimal_2021}. 

    \hspace{0.5cm}$\bullet$ Case $\rho(0)= 0$. It is not possible to reset very close to the target site and Eq. (\ref{Trcase1}) is not valid. An example is when $\rho(z)$ vanishes everywhere except in a compact support, i.e., $\rho(z)\neq0$ only in some interval $[z_{min},z_{max}]$ with $0<z_{min}<z_{max}<\infty$. At large $r$, the exponential factor in Eq. (\ref{pdf-4}) decays rapidly with $z$. If  $\rho(z)$ varies smoothly for $z>z_{min}$, it can be replaced by $\rho(z_{min})$ over the whole interval, provided $\rho(z_{min})\neq0$. Under these assumptions, Eq. (\ref{pdf-4}) reads
\begin{equation}
\mathcal{T}_{r}\simeq \frac{e^{\alpha_0z_{min}}}{\rho(z_{min})\sqrt{Dr}}\to\infty .
\label{Trcase2}
\end{equation}
    From Eqs. (\ref{smallr}) and (\ref{Trcase2}), one expects that $\mathcal{T}_r$ has at least one local minimum and that $r^*$ is nontrivial.

In the following, we analyze Eq. \eqref{pdf-4} for choices of $\rho(z)$ belonging to the second case $\rho(0)=0$ above. In Ref. \cite{two_resetting_points} we considered a particular example exhibiting a rich phenomenology, that is, two resetting points at positions $z_1$ and $z_2$. In that case, $\rho(z)= a_1\delta(z-z_1)+a_2\delta(z-z_2)$ with $0<z_1<z_2$ and $a_1+a_2=1$ by normalization. We showed that the function ${\cal T}_r$ of $r$ could exhibit two local minima (instead of one) while the optimal rate $r^*$ corresponding to the global minimum underwent a discontinuous \lq\lq first-order" transition as $z_2$ was varied (fixing $z_1=1$). This discontinuous jump existed provided that $a_2/a_1$ was large enough and exceeded a non-trivial critical value of $6.600817\ldots$ Below this threshold, the variations of $r^*$ with $z_2$ were smooth, although not necessarily monotonously decreasing like in single-site resetting \cite{two_resetting_points}. 

\section{First-order transitions in the optimal resetting rate}
\label{sec:other_rho}
We consider several resetting point distributions with compact support and that are symmetric around their mean $\mu$. We use dimensionless space and time units. 
By definition, the left edge of the support is located at a unit distance from the target at the origin, and the diffusion coefficient is set to $D=1$. The upper limit of the interval is $z_{max} = 2\mu - 1$, so that the middle point of $[1,2\mu - 1]$ coincides with the average resetting position at $z = \mu$, with $\mu> 1$. In summary, the distribution $\rho(z)$ is non-zero in the interval $[1,2\mu-1]$ and vanishes everywhere else, as depicted in Fig. \ref{fig:illustration}.
Two simple choices of $\rho(z)$ are the truncated Gaussian and truncated exponential (Laplace) distributions.
    
\subsection{Truncated Gaussian Distribution}
As a first example, let us consider the truncated Gaussian distribution (TGD) with mean $\mu>1$ and variance parameter $\sigma^{2}$
\begin{equation}
	\rho \left( z\right) =\frac{e^{-\left( z-\mu \right) ^{2}/2\sigma ^{2}}}{%
		\sqrt{2\pi \sigma ^{2}}\erf\left( \frac{\mu -1}{\sqrt{2\sigma ^{2}}}\right) }%
	,\quad 1\leq z\leq 2\mu -1,\label{pdf-5}
\end{equation}
while $\rho(z) = 0$ for $z<1$ and $z>2\mu-1$. The normalization factor in Eq. (\ref{pdf-5}) ensures that $\int_{1}^{2\mu -1}\rho(z) \dd z = 1$. The uniform distribution is recovered in the limit $\sigma\to\infty$.
After some manipulations, the substitution of Eq. \eqref{pdf-5} into Eq. \eqref{pdf-4} gives
\begin{equation}
	\mathcal{T}_{r}\left( \mu ,\sigma \right) =\frac{1}{r}\left( \frac{%
		2e^{-\alpha _{0}^{2}\sigma ^{2}/2}e^{\alpha _{0}\mu }\erf\left( \frac{\mu -1%
		}{\sqrt{2\sigma ^{2}}}\right) }{\erf\left( \frac{\mu -1+\alpha _{0}\sigma
			^{2}}{\sqrt{2}\sigma }\right) -\erf\left( \frac{1-\mu +\alpha _{0}\sigma ^{2}%
		}{\sqrt{2}\sigma }\right) }-1\right),  \label{pdf-6}
\end{equation}
with $\alpha_0$ given by Eq. (\ref{alpha0}).
\begin{figure}
\centering
\includegraphics[width=\linewidth]{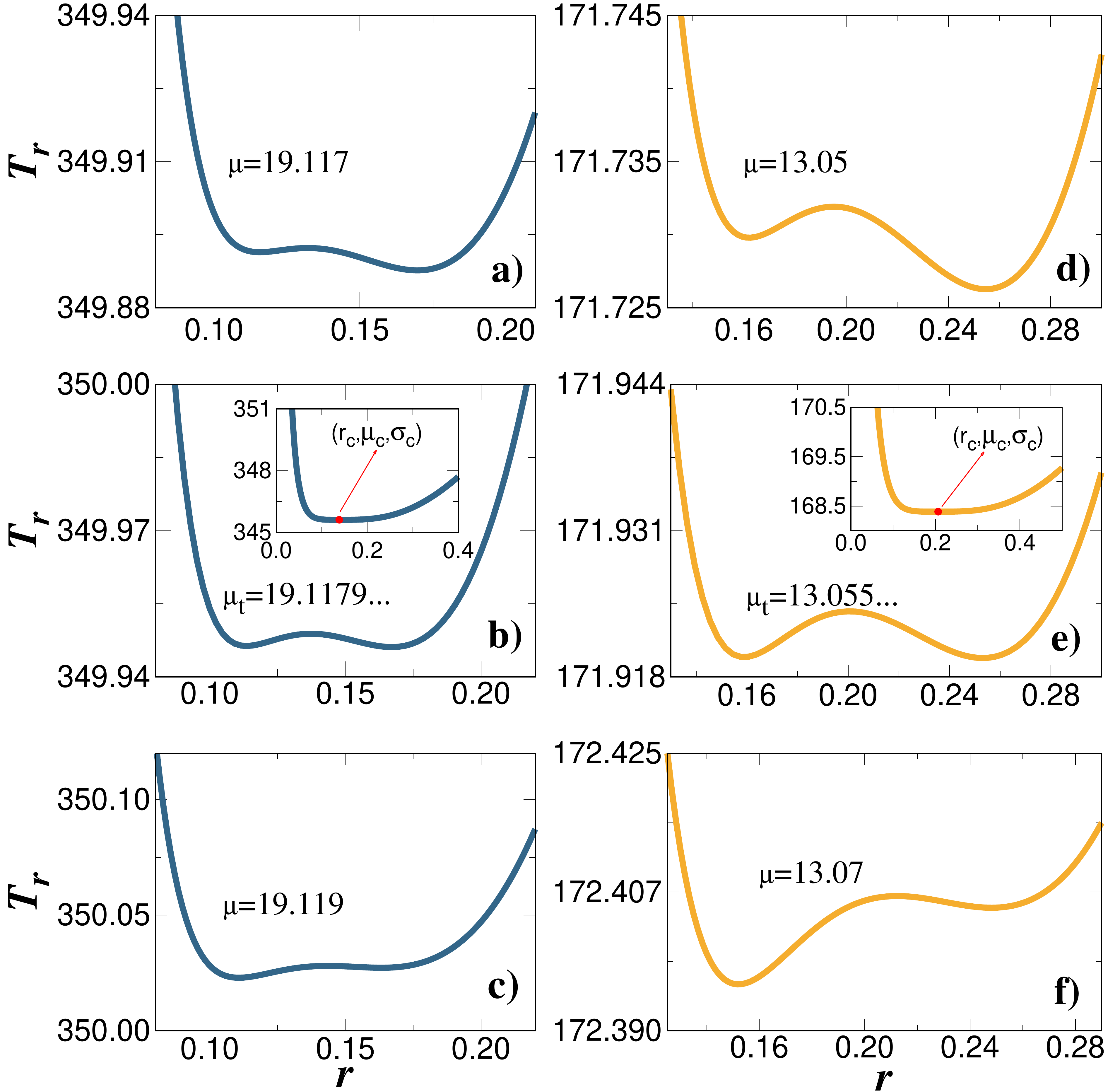}
  \caption{Left column: averaged mean first-passage time ${\cal T}_r$ given by Eq. \eqref{pdf-6} as a function of $r$, when the resetting distribution is a truncated Gaussian with $\sigma=8.8$ and: (a) $\mu=19.117$, (b) $\mu=\mu_{t}=19.1179...$, (c) $\mu=19.119$. Right column: Same quantity in the truncated exponential case, see Eq. \eqref{pdf-8}, fixing $\sigma=8.26$ and: (d) $\mu=13.05$, (e) $\mu=\mu_{t}=13.0557...$, (f) $\mu=13.07$. 
  The insets in panels (b) and (e) show ${\cal T}_r$ at the critical parameters $\sigma_{c}$ and $\mu_c$ where the two minima merge into a single one, located at $r=r_c$. The critical parameter values are obtained further in Section \ref{sec:GL} by solving Eqs. \eqref{pdf-9}. One obtains $(\mu_c,\sigma_c,r_c)\simeq(19.0105,8.7610,0.1382)$ and $(12.9373,8.2355,0.2028)$ for the TGD and TED, respectively. For $\sigma < \sigma_{c}$, ${\cal T}_r$ admits a single minimum (not shown). }\label{fig:mfpt_gauss_expo}
\end{figure}
%
    As ${\cal T}_r$ always diverges at small and large $r$ for this kind of distribution, see Eqs. (\ref{smallr}) and (\ref{Trcase2}), there should exist at least one local minimum of $\mathcal{T}_{r}$ for each pair $\left(
    \mu ,\sigma \right)$.

As an illustrative example, in the left column of Fig. \ref{fig:mfpt_gauss_expo} we fix $\sigma=8.8$ and plot $\mathcal{T}_{r}$ as a function of $r$ for different values of $\mu$. In Fig. \ref{fig:mfpt_gauss_expo}a, where $\mu=19.1170$, the AMFPT presents two local minima, at $r_1$ and $r_2$ (with $r_2<r_1$), and the global minimum corresponds to the larger value $r_1$. In Fig. \ref{fig:mfpt_gauss_expo}b, at a slightly larger $\mu=\mu _{t}=19.1179...$, a transition point is reached where the two local minima denoted as $r_1^*$ and $r_2^*(<r_1^*)$ are equivalent, i.e., ${\cal T}_{r_1^*}={\cal T}_{r_2^*}$, hence the two resetting rates $r_1^*$ and $r_2^*$ are equally optimal. Note that $\mu_t$, $r_1^*$ and $r_2^*$ depend on $\sigma$, {\em a priori}. Finally, as we see from Fig. \ref{fig:mfpt_gauss_expo}c, if $\mu$ further increases from $\mu_t$, the previous global minimum becomes secondary, i.e., the optimal rate $r^{*}$ drops abruptly from $r_1^*$ to $r_2^*$. 

This discontinuous transition scenario is observed only if $\sigma$ is large enough, that is, greater than a critical value given by $\sigma_c=8.761047...$. For $\sigma<\sigma_c$, the AMFPT ${\cal T}_r$ versus $r$ admits only one minimum for all $\mu$, and $r^*$ varies smoothly with $\mu$. When $\sigma$ approaches $\sigma_c$ from above, the two equivalent minima of Fig. \ref{fig:mfpt_gauss_expo}b tend to merge into a single minimum, i.e., $r_1^*$ tends to $r_2^*$, while the transition value $\mu_t$ tends to $\mu_c=19.0105...$. In the inset, we have represented ${\cal T}_r$ vs. $r$ at $\sigma=\sigma_c$ and $\mu=\mu_c$. The position of the single minimum defines a critical rate $r_1^*=r_2^*=r_c$, which takes the value $r_c=0.13829...$ These precise values of the critical parameters are obtained from a numerical calculation presented in Section \ref{sec:GL} below. In parameter space, the triplet $(r_c,\mu_c,\sigma_c)$ defines a critical point analogous to a liquid-gas critical point, as we discuss further.

\begin{figure*}
  \centering
    \includegraphics[width=0.48\linewidth]{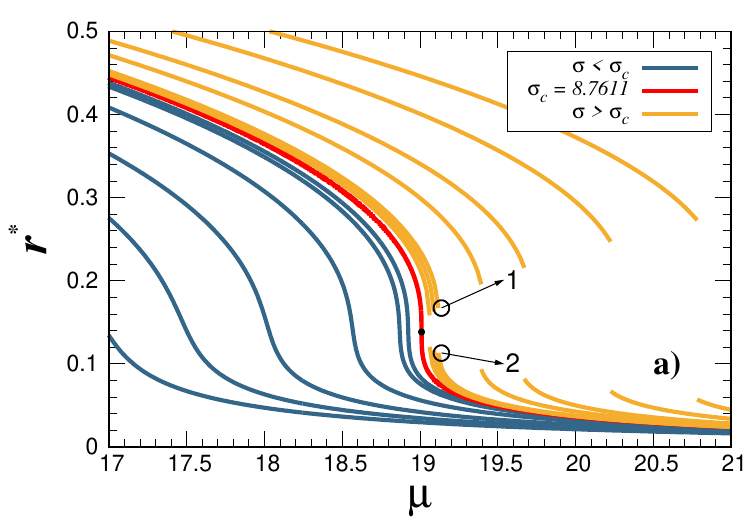}
    \label{fig:jumps_r_T_a}
    \includegraphics[width=0.48\linewidth]{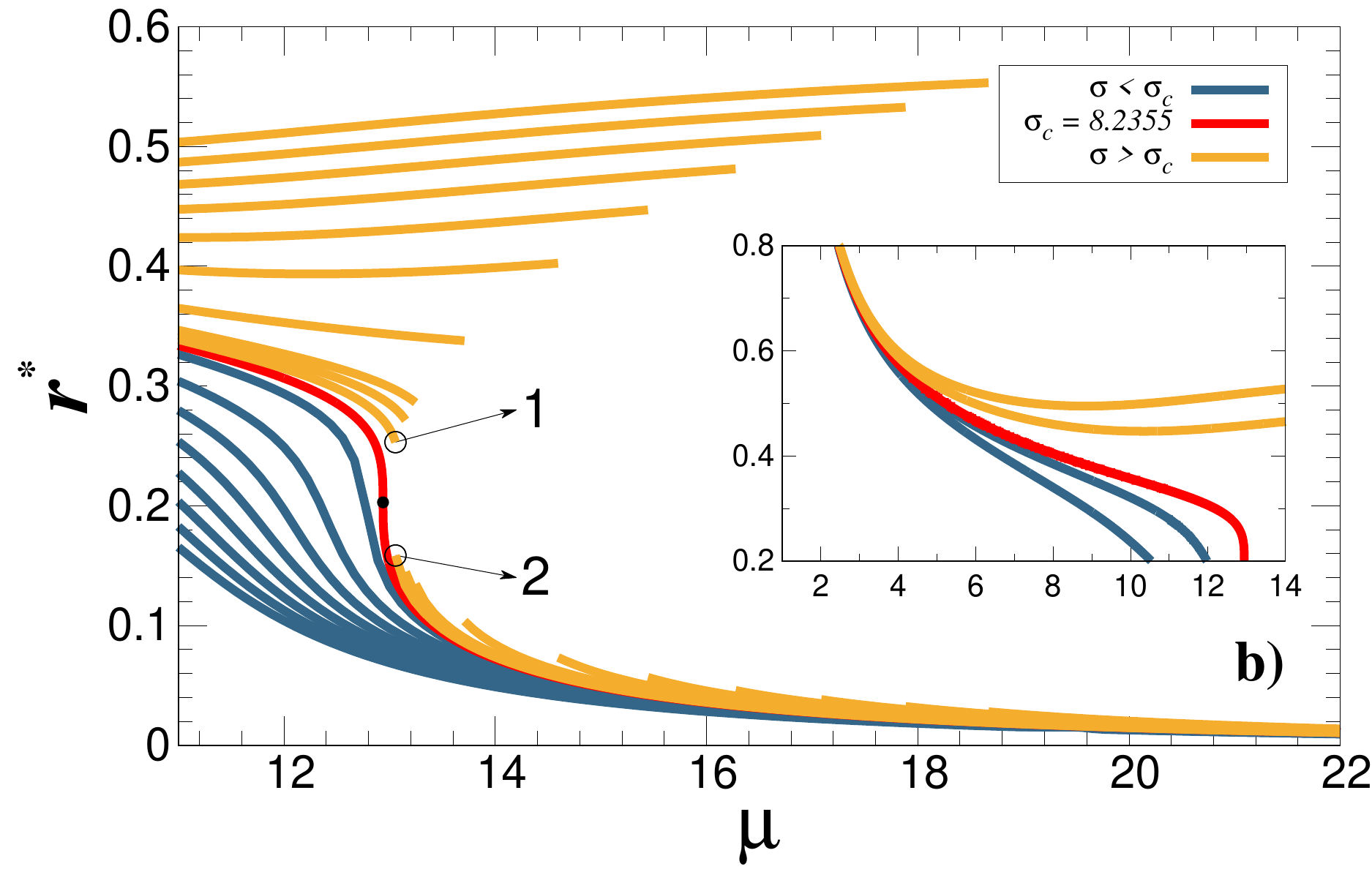}
    \label{fig:jumps_r_T_b}
  \caption{
Optimal resetting rate $r^{*}$ minimizing Eq. \eqref{pdf-6} in the TGD case [panel (a)], and Eq. \eqref{pdf-8} in the TED case [panel (b)], as a function of $\mu$ for various $\sigma$. The critical point with diverging derivative is shown in black in each case. For any value of $\sigma$ greater than $\sigma_c$, i.e., to the right of the critical curve in red, $r^*$ exhibits a discontinuity at  $\mu=\mu_t(\sigma)$. The labels 1 and 2 denote the minima in Figs. \ref{fig:mfpt_gauss_expo}b and \ref{fig:mfpt_gauss_expo}e at the transition. The inset of (b) is a zoom of the small $\mu$ region that illustrates the non-monotonous character of $r^*$ when $\sigma>\sigma_c$.} \label{fig:jumps_r_T}
\end{figure*}

The results can be summarized by representing the optimal resetting rate $r^{\ast}$ as a function of $\mu$ and $\sigma$. The optimal rate is obtained numerically from its definition,  
\begin{equation}
r^{*}(\mu,\sigma) = \underset{r\ge 0}{\arg \min}\,\mathcal{T}_r(\mu,\sigma), \label{pdf-6A}
\end{equation}
and displayed in Fig. \ref{fig:jumps_r_T}a as a function of $\mu$ for various $\sigma$.
One can appreciate three kinds of curves, all of them displaying a monotonous decreasing behavior of $r^{*}$ with the average resetting position $\mu$. 

For $\sigma < \sigma _{c}$ (blue curves), the optimal rate varies smoothly with $\mu$ throughout the interval $1 < \mu < \infty$. This behavior is a consequence of the unique local minimum of ${\cal T}_r$ at $r^*$, and qualitatively resembles the limiting case of resetting to a single point, which corresponds to $\sigma=0$ or $\rho(z)=\delta(z-\mu)$ \cite{evans_diffusion_2011,evans_diffusion_2011-1,evans_stochastic_2020,evans2014diffusion}. Nevertheless, as $\sigma$ increases and approaches the non-trivial value $\sigma_c$, the slope of the blue curves becomes progressively steeper at a particular value of $\mu$.

When $\sigma= \sigma_{c} = 8.761047...$  (red curve), there exists a value of $\mu$, located at $\mu_c = 19.0105...$, which is singular: at this point, $\partial_{\mu} r^*|_{\mu=\mu_c}=-\infty$. Hence, the curve $\sigma=\sigma_c$ is \lq\lq critical", in the sense that it contains a singularity with coordinates $(\mu_c,r_c)$ shown with a black dot in Fig. \ref{fig:jumps_r_T}a. This point corresponds to the one discussed above in the inset in Fig. \ref{fig:mfpt_gauss_expo}b. 

For $\sigma > \sigma_{c}$ (yellow curves), the optimal resetting rate $r^{\ast}$ exhibits a discontinuity at a transition value $\mu=\mu_t(\sigma)>\mu_c$. 
The points encircled and labeled by 1 and 2 in Fig. \ref{fig:jumps_r_T}a correspond to the equivalent minima at $r_1^*$ and $r_2^*$ of Fig. \ref{fig:mfpt_gauss_expo}b. At each transition, two different optimal resetting rates thus \lq\lq coexist". 
In other words, for a sufficiently wide Gaussian in Eq. \eqref{pdf-5} and with a mean sufficiently far from the target ($\mu > \mu_{c}$), it is possible to minimize the mean search time with two different resetting rates. These coexisting optimal rates tend to differ widely as $\sigma$ increases; see Fig. \ref{fig:jumps_r_T}a. 
Note that a similar phenomenology holds
for the case of two resetting points \cite{two_resetting_points}, with some important differences to be mentioned later.

\subsection{Truncated exponential distribution}
Another example is the symmetric, truncated exponential
distribution (TED) centered at $\mu$ and vanishing outside the interval $[1,2\mu -1]$,
see Fig. \ref{fig:illustration}c. As for the Gaussian, the parameter $\sigma$ denotes the standard deviation of the distribution if it was not truncated, hence we write $\rho(z)$ under the form 
\begin{equation}
\rho \left( z\right) =\frac{e^{-\frac{\sqrt{2}\left\vert z-\mu \right\vert }{%
\sigma }}}{\sqrt{2}\sigma \left[ 1-e^{-\frac{\sqrt{2}}{\sigma }\left( \mu
-1\right) }\right] },\text{ \ \ }1\leq z\leq 2\mu -1,  \label{pdf-7}
\end{equation}%
so that it is normalized. After substitution of Eq. \eqref{pdf-7} into Eq. \eqref{pdf-4}, one obtains the AMFPT to be minimized with respect to $r$,
\begin{widetext}
\begin{equation}
\mathcal{T}_{r}(\mu,\sigma)=\frac{1}{r}\left( \frac{\sqrt{2}\left( 2-\alpha _{0}^{2}\sigma
^{2}\right) \left( 1-e^{-\frac{\sqrt{2}\left( \mu -1\right) }{\sigma }
}\right) e^{\mu \alpha _{0}}}{2\sqrt{2}+\left( \alpha _{0}\sigma -\sqrt{2}
\right) e^{-\left( \mu -1\right) \left( \alpha _{0}+\frac{\sqrt{2}}{\sigma }
\right) }-\left( \alpha _{0}\sigma +\sqrt{2}\right) e^{\left( \mu -1\right)
\left( \alpha _{0}-\frac{\sqrt{2}}{\sigma }\right) }}-1\right).
\label{pdf-8}
\end{equation}
\end{widetext}

The results obtained from evaluating Eq. \eqref{pdf-8} are qualitatively similar to the Gaussian case, as shown
in the right column of Fig. \ref{fig:mfpt_gauss_expo}. In this example, one fixes $\sigma=8.26$ and $\mu$ takes the values $13.05$, $13.0557\simeq\mu_t$ and $13.07$, in panels (d),(e) and (f), respectively. The global minimum of ${\cal T}_r$ with respect to $r$ exhibits a discontinuity at $\mu_t$.
For the TED case, crossing $\mu_t$ from below also results in a negative jump of $r^*$ (from $r_1^*$ to $r_2^*<r_1^*$).
Once again, the inset of Fig. \ref{fig:mfpt_gauss_expo}e displays Eq. \eqref{pdf-8} as a function of $r$ for the critical mean and variance $\left(\mu_{c}, \sigma_{c}\right)= \left(12.9376..., 8.2355...\right)$, where the minimum is reached at a single point $r_{c} = 0.2028...$. For $\sigma<\sigma_c$, ${\cal T}_r$ displays a single minimum.

The optimal rate obtained from numerical minimization is shown in Fig. \ref{fig:jumps_r_T}b. Again, the equivalent minima corresponding to Fig. \ref{fig:mfpt_gauss_expo}e are highlighted by labels 1 and 2. Although $r^*\to 0$ when $\mu\to\infty$, Fig. \ref{fig:jumps_r_T}b and its inset show that $r^*$ is not always monotonously decreasing, unlike the TGD case. At large enough $\sigma$, the optimal rate can increase with $\mu$ in a certain intermediate interval, before the abrupt jump from $r_1^*$ to $r_2^*$. 

This non-intuitive behavior is quite subtle and probably due to the fact that the exponential function decays to 0 not as fast as the Gaussian. The qualitative scenario suggested by Fig. \ref{fig:jumps_r_T}b (when $\sigma>\sigma_c$) is the following. At small $\mu$, the support is narrow and the distribution $\rho(z)$ nearly uniform in that support. Since all the resetting positions are relatively close to the target, the optimal rate is high. As the mean resetting position $\mu$ increases, the distribution is still nearly uniform but widens, and the optimal resetting rate decreases (like in the TGD case). At some large enough $\mu$, however, the diffusive paths that start from the central part of $\rho(z)$ take a long time to reach the target. Meanwhile, the probability density at the left-edge, $\rho(z=1) \sim e^{-(\mu-1)/(\sqrt{2}\sigma)}$, is still not so small. It is thus convenient to increase the resetting rate, so that the time interval between successive resetting events will favor the success of diffusive paths that start close to the target, which is an advantageous region. Consequently, $r^*$ increases. As $\mu$ further increases, $\rho(z=1)$ becomes too small: resetting close to the target occurs too rarely. Hence the best strategy is to adjust the resetting rate to the most likely resetting sites, located near $\mu$, far from the target, causing a jump of $r^*$ to a much smaller value.

The existence of a critical point and the similar behavior of $r^{*}$ in its vicinity for the TGD and TED cases suggest that a general phenomenological theory can be developed and applied to a wide class of distributions $\rho(z)$.

\section{Phenomenological Ginzburg-Landau theory}\label{sec:GL}

The purpose of this Section is to write a Taylor expansion of ${\cal T}_r$ for $r$ close to $r_c$ that captures the different shapes observed in the panels and insets of Fig. \ref{fig:mfpt_gauss_expo}, when the parameters $\mu$ and $\sigma$  are varied around $\mu_c$ and $\sigma_c$, respectively. 

We first need to determine the values of the critical parameters, when they exist. This critical point is characterized by a change in the concavity of ${\cal T}_r$ at $r=r_{c}$, as $r_c$ is a unique minimum of ${\cal T}_r$ for $\sigma\le\sigma_c$ and a local maximum for $\sigma>\sigma_c$ (in any case, it is assumed that $\sigma$ is close to $\sigma_c$). This implies vanishing first and second derivatives at this point. For $\sigma>\sigma_c$ but close to $\sigma_c$, if one sets $\mu=\mu_t(\sigma)\simeq\mu_c$, one finds two symmetric, equivalent minima at $r_1^*(\sigma)$ and $r_2^*(\sigma)$ on each side of the local maximum  $r_c$, with $r_2^*(\sigma)<r_c<r_1^*(\sigma)$. Hence, the Taylor expansion of ${\cal T}_r$ should not include a cubic term $(r-r_c)^3$ (although asymmetry can become important further away from the critical point; see, e.g., Fig. \ref{fig:prob_vs_xr_Tvsr}a) further. One deduce from the above considerations that the first three derivatives of ${\cal T}_r$ with respect to $r$ must vanish at the critical point,
\begin{equation}
	\left.\frac{\partial^n \mathcal{T}(r,\mu,\sigma)}{\partial r^n} \right\vert_{\left(
			r,\mu,\sigma\right) =\left(
			r_{c},\mu_{c},\sigma_{c} \right) } =0\quad {\rm for}\ n=1,2,3.
	\label{pdf-9}
\end{equation}
The equation with $n=1$ follows from the fact that $r=r_c$ is a minimum by definition (optimal rate) at $\sigma=\sigma_c$ and $\mu=\mu_c$, as illustrated by the insets of Figs. \ref{fig:mfpt_gauss_expo}b and \ref{fig:mfpt_gauss_expo}e. Relations analogous to Eqs. (\ref{pdf-9}) were used in \cite{two_resetting_points} for the two-point problem.
Hence the critical parameters for the TGD and TED cases can be calculated by numerically solving the system of equations (\ref{pdf-9}) with the help of Mathematica, using Eqs. \eqref{pdf-6} and \eqref{pdf-8}, respectively. For the TGD, one obtains the unique solution,
\begin{eqnarray}
&&\mu_c= 19.0105...\nonumber\\
&&\sigma_c=8.7610...\\
&& r_c=0.1382... ,\nonumber
\end{eqnarray}
while for the TED,
\begin{eqnarray}
&&\mu_c= 12.9376...\nonumber\\
&&\sigma_c=8.2355...\\
&& r_c=0.2028...\ .\nonumber
\end{eqnarray}
These are the values reported previously in Section \ref{sec:other_rho}. 
Note that for other choices of $\rho(z)$, Eqs. (\ref{pdf-9}) may not have a solution. Examples are the classical single-point resetting problem, with $\rho(z)=\delta(z-\mu)$, or the truncated uniform distribution, with $\rho(z)=1/(2\mu-2)$ for $1\le z\le 2\mu-1$ \cite{mendez_first_passage}. In these cases, ${\cal T}_r$ has a unique local minimum at $r=r^*(\mu)$ for all $\mu$, and the second derivative remains strictly positive at this point.

Assuming that a critical point exists, one deduces from Eq. (\ref{pdf-9}) that the simplest form of ${\cal T}_r$ at criticality is
\begin{equation}
{\cal T}_r\simeq  T_c+\frac{a_4}{4}(r-r_c)^4 ,\quad\mu=\mu_c,\ \sigma=\sigma_c,
\label{gl-1}
\end{equation}
with $T_c$ a constant time and $a_4$ a positive coefficient depending little on $(\mu,\sigma)$ near $(\mu_c,\sigma_c)$. This behavior describes the inset in Fig. \ref{fig:mfpt_gauss_expo}b or \ref{fig:mfpt_gauss_expo}e. If $\sigma>\sigma_c$ and $\mu=\mu_t(\sigma)$, the quadratic term is not zero (negative) but ${\cal T}_r$ is still symmetric in the vicinity of $r_c$ (see, e.g., Fig. \ref{fig:mfpt_gauss_expo}b). Eq. (\ref{gl-1}) becomes
\begin{eqnarray}
{\cal T}_r&\simeq&  T_c+ \frac{a_2}{2}(r-r_c)^2+\frac{a_4}{4}(r-r_c)^4\, ,\label{gl-2}\\
&&\quad\mu=\mu_t(\sigma),\ \sigma\ge \sigma_c,\nonumber
\end{eqnarray}
where $a_2<0$. The simplest choice is
\begin{equation}
a_2=-\alpha_2(\sigma-\sigma_c),\quad {\rm with}\ \alpha_2>0. 
\end{equation}
Eq. (\ref{gl-2}) holds for $\sigma<\sigma_c$ and $\mu\simeq\mu_c$ as well: in this case $a_2>0$ and ${\cal T}_r$ has a unique minimum at $r^*=r_c$.
In the more general case where $\sigma>\sigma_c$ and $\mu\neq\mu_t(\sigma)$, one of the two local minima at $r_1$ and $r_2$ is favored and ${\cal T}_r$ is no longer symmetric around $r_c$ (see, e.g., Fig. \ref{fig:mfpt_gauss_expo}a or \ref{fig:mfpt_gauss_expo}c). Therefore, a linear term in $r-r_c$ must be added to Eq. (\ref{gl-2}),
\begin{eqnarray}
{\cal T}_r&\simeq&  T_c+a_1(r-r_c)+\frac{a_2}{2}(r-r_c)^2+\frac{a_4}{4}(r-r_c)^4,\label{gl-3}\\ 
&&\quad \mu\neq\mu_t(\sigma),\ \sigma\ge\sigma_c.\nonumber
\end{eqnarray}
The simplest form for $a_1$ is
\begin{equation}
\label{a1tgd}
a_1=\alpha_1(\mu-\mu_t(\sigma)),\quad {\rm with}\ \alpha_1>0.
\end{equation}
The coefficient $\alpha_1$ must be positive because we wish the minimum at the larger rate $r_1>r_c$ to be the global minimum of ${\cal T}_r$ when $\mu<\mu_t(\sigma)$ (see, e.g., Fig. \ref{fig:mfpt_gauss_expo}d). In this case $a_1<0$ and the linear term in Eq. (\ref{gl-3}) will lower ${\cal T}_r$ for $r>r_c$. In a ferromagnetic analogy, this is comparable to having a positive external magnetic field that favors the spin-up configuration. 
Given the complexity of the expressions (\ref{pdf-6}) and (\ref{pdf-8}), we have not attempted to explicitly compute $\alpha_1$, $\alpha_2$ and $a_4$ for the TGD and TED.

The minimization of Eq. (\ref{gl-3}) with respect to $r$ gives
\begin{equation}
    \alpha_{1}(\mu - \mu_{t}(\sigma)) - \alpha_{2}(\sigma - \sigma_c)(r^{\ast} - r_{c}) + a_{4}(r^{*}-r_{c})^{3} = 0,
    \label{mini}
\end{equation}
from which one deduces two scaling laws. By setting $\sigma=\sigma_c$ in Eq. (\ref{mini}), one obtains an equation for the critical line representing the variations of $r^*$ with $\mu$ near the singular point of Figs. \ref{fig:jumps_r_T}a-b,
\begin{equation}
    r^{*}(\mu)\vert_{\sigma=\sigma_c} \simeq r_{c} + \mathcal{B}\vert \mu_{c} - \mu \vert ^{1/\delta}{\rm sign}(\mu_{c}-\mu). \label{pdf-11}
\end{equation}
The critical exponent is $\delta = 3$ and $\mathcal{B} = \sqrt[3]{\alpha_1/a_4}$. On the other hand, setting $\mu=\mu_t(\sigma)$ in Eq. (\ref{mini}) one obtains the behavior of the \lq\lq order parameter", the size of the discontinuity $\Delta r^*(\sigma)\equiv r_1^*(\sigma)-r_2^*(\sigma)$ at the transition when $\sigma-\sigma_c$ is positive and small,
\begin{equation}
    \Delta r^{*}(\sigma) \simeq \mathcal{B}'(\sigma - \sigma_{c})^{\beta}. \label{pdf-10}
\end{equation}
The second critical exponent is $\beta=1/2$ and $\mathcal{B}' = 2\sqrt{\alpha_2/a_4}$.

\section{Last resetting position before absorption}\label{sec:lastresetting}
In the presence of multiple resetting points, we can explore a question specific to distributed resetting: in what position the particle was reset just before being absorbed by the target? Let us introduce the PDF of the last resetting point, $\rho_l(z)$, defined as the probability that the particle was reset near $z$ at some time, given that it subsequently diffused from there and was absorbed without further resetting. Since the last resetting position $z$ is conditioned to further absorption, $\rho_l(z)\neq\rho(z)$.
This quantity gives information on how useful is a given position $z$ for the search process. In particular, the region of space where $\rho_l(z)$ is larger indicates those resetting positions that are more efficient for a given choice of $\rho(z)$ and $r$.

Let us consider an arbitrary underlying process (with $r=0$) for which the PDF of the first passage time $t$ at the origin is known and denoted as ${\cal P}_0(t|z)$, where $z$ is the initial position. For the process subject to resetting, let us introduce the joint probability density function ${\cal P}_{r}(t,z)$ of the first passage time $t$ at the origin and of the position $z$ where the underlying process was reset for the last time. We have the general relation,
\begin{multline}
    \mathcal{P}_{r}(t,z) = \rho(z)e^{-rt}\mathcal{P}_{0}(t|z)\\ + r\rho(z)\int_{-\infty}^{\infty}\rho(x_0)\dd x_{0}\int_{0}^{t}e^{-r\tau}S_{r}(t-\tau|x_0)\mathcal{P}_{0}(\tau|z)\dd \tau. \label{pdf-14}
\end{multline}
The construction of $\mathcal{P}_{r}$ is similar to Eq. \eqref{pdf-1}. The first term of the rhs of Eq. (\ref{pdf-14}) takes into account the contribution from the trajectories that started near $z$ at $t=0$ and found the target at time $t$, without any resetting. The second term represents those trajectories that started from some position $x_0$, underwent at least one reset and had their last resetting at time $t-\tau$ near $z$ before being absorbed $\tau$ time units later.  
The last resetting point density follows from the marginal $\rho_{l}(z) = \int_{0}^{\infty} \mathcal{P}_{r}(t,z)\dd t$. Integrating Eq. \eqref{pdf-14} over $t$ gives after straightforward algebra
\begin{equation}
    \rho_{l}(z) =(r \mathcal{T}_{r} +1) \rho(z) \widetilde{{\cal P}}_0(r|z), \label{pdf-16}
\end{equation}
where $\widetilde{{\cal P}}_0(s|z)$ is the Laplace transform of ${\cal P}_0(t|z)$ and 
where we have used Eq. (\ref{amfpt}) with the general identity  $T(x_0)=\int_{0}^{\infty} \dd t S_{r}(t|x_0)$. 
For a Brownian particle, ${\cal P}_0(t|z)$ is given by the L\'evy-Smirnov distribution \cite{dagdug_diffusion_2024}
\begin{equation}
    \mathcal{P}_{0}(t|z) = \frac{|z|}{\sqrt{4\pi D}}\frac{e^{-z^{2}/4Dt}}{t^{3/2}},\label{pdf-13}
\end{equation}
whose Laplace transform is $\widetilde{{\cal P}}_0(s|z)=e^{-\sqrt{s/D}|z|}$. One can check using Eq. (\ref{pdf-4}) that $\rho_l(z)$ in Eq. (\ref{pdf-16}) is normalized. The sought distribution thus reads 
\begin{equation}
\rho_{l}(z) =\mathcal{C}\rho(z)e^{-\alpha_{0}|z|} , \quad \alpha_{0} = \sqrt{\frac{r}{D}}
\label{pdf-17},
\end{equation}
with $\mathcal{C}$ the normalization constant. 

The expression (\ref{pdf-17}), or more generally (\ref{pdf-16}), is one of our main result and reveal an important aspect of distributed resetting: $\rho_l(z)$ is a product of two terms that can have antagonistic effects. For a given $z$, the density $\rho_l(z)$ can be high for essentially two reasons: 

\hspace{0.5cm}$\bullet$ $\rho(z)$ is large, i.e., $z$ is frequently used as a resetting position and therefore likely to be the last one; 

\hspace{0.5cm}$\bullet$ $z$ is close to the target ($|z|$ small), making the exponential factor in Eq. (\ref{pdf-17}) not so small. Hence $z$ may not be a frequently used position, but from this resetting point the capture probability is high. 

Loosely speaking, $\rho_l(z)$ in Eq. (\ref{pdf-17}), is analogous to a Boltzmann-Gibbs distribution of energies, where $|z|$ would stand for the energy and $\rho(z)$ for the number of states with energy $|z|$. In the first case above \lq\lq entropic" effects dominate (frequent positions are useful), whereas in the second one \lq\lq energetic" effects dominate (positions close to the target are useful). The competition between these two trends gives rise to rich behaviors, and the discontinuous transitions can be re-analyzed under this scope. 

\subsection{Truncated distributions}

With the truncated Gaussian in Eq. (\ref{pdf-5}), the last resetting point PDF $\rho_l(z)$ is also a truncated Gaussian. Its explicit expression for $1\le z\le 2\mu-1$ is
\begin{equation}
    \rho_{l}(z) = \frac{%
		2e^{-\frac{\alpha _{0}^{2}\sigma ^{2}}{2}}e^{\alpha _{0}\mu } e^{-\left( z-\mu \right) ^{2}/2\sigma ^{2}} e^{-\alpha_{0}z}}{\sqrt{2\pi\sigma^{2}}\left[\erf\left( \frac{\mu -1+\alpha _{0}\sigma
			^{2}}{\sqrt{2}\sigma }\right) -\erf\left( \frac{1-\mu +\alpha _{0}\sigma ^{2}%
		}{\sqrt{2}\sigma }\right) \right ]}, \label{pdf-22}
\end{equation}
and its most probable position $\hat{z}_l\equiv{\arg \max}_{z}\,\rho_{l}(z)$ is shifted to the left of $\mu$ and given by $\hat{z}_l={\rm max[1,\mu-\alpha_0\sigma^2]}$. In other words,
\begin{equation}
\hat{z}_l=
\begin{cases}
\mu-\sqrt{\frac{r}{D}}\sigma^2\quad {\rm if}\ r <D\left(\frac{\mu-1}{\sigma^2}\right)^2\\
\\
1\quad {\rm if}\ r >D\left(\frac{\mu-1}{\sigma^2}\right)^2.
\end{cases}
\label{hatz}
\end{equation}
Given a distribution $\rho(z)$ with parameters $(\mu,\sigma)$, Eq. (\ref{hatz}) shows that for $r$ large enough, $\rho_l(z)$ is peaked at $z=1$, the left edge of the support. Hence, successful diffusive paths are those that tend to start close to the target (\lq\lq energetic" regime). Conversely, for very small $r$, $\rho_l(z)\simeq\rho(z)$ hence the most successful paths are those that start where $\rho(z)$ is maximal (\lq\lq entropic" regime).

The optimal rate can belong to either qualitative regime, depending on $(\mu,\sigma)$. The large discontinuities in $r^*$ in Fig. \ref{fig:jumps_r_T}a or \ref{fig:jumps_r_T}b, can actually be interpreted as an abrupt regime change. Fixing $\sigma$, before the discontinuity ($\mu<\mu_t(\sigma)$), the optimal rate is relatively large, indicating that this strategy relies on the positions that are closer to the target. However, after the discontinuity ($\mu>\mu_t(\sigma)$), the rate can be significantly smaller, so that the searcher is optimal thanks to the central part of $\rho(z)$, which is farther away but where resetting predominantly occurs. This regime change can be understood by noticing that when $\mu$ increases at fixed $\sigma$, it becomes very unlikely to reset near the left edge, since $\rho(z=1)\sim e^{-\mu^2/2\sigma^2}\to 0$. Therefore, the optimal resetting rate must be small in order to allow the paths starting from the central part of the distribution to reach the target.

\begin{figure}
\centering
\includegraphics[width=\linewidth]{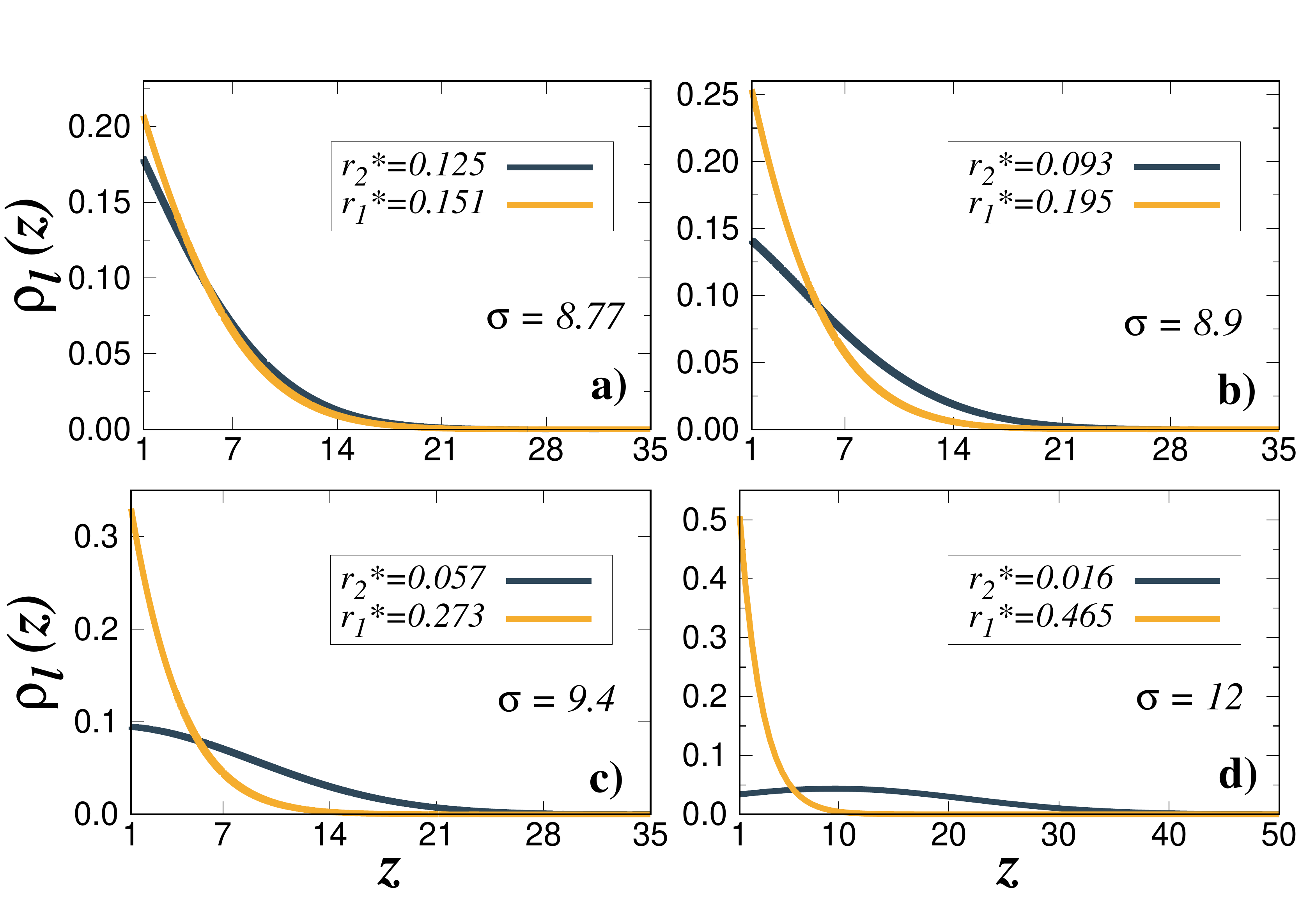}
  \caption{Last resetting point PDF given by Eq. \eqref{pdf-22} at the two equivalent optimal rates $r_1^*(\sigma)$ and $r_2^*(\sigma)$, for different values of $\sigma > \sigma_{c}$. The corresponding transition values $\mu_t(\sigma)$ are $19.0352$, $19.3941$, $20.7817$, and $28.1604$, in panels (a), (b), (c) and (d) respectively.}\label{fig:rho_l_vs_z}
\end{figure}

In Fig. \ref{fig:rho_l_vs_z}, we verify these predictions by showing the distributions $\rho_l(z)$ corresponding to the two coexisting optimal rates $r_1^*(\sigma)$ and $r_2^*(\sigma)$. 
Clearly, near the critical point ($\sigma \approx \sigma_c$) the two curves in Fig. \ref{fig:rho_l_vs_z}a are peaked at $z=1$ and rapidly vanish in the bulk. They thus belong to the \lq\lq energetic" regime and are essentially superposed because the two optimal rates are very close to each other. However, as $\sigma-\sigma_c$ grows, the difference between $r_1^*(\sigma)$ and $r_2^*(\sigma)$ becomes more evident. While $\rho_l(z)$ for $r=r_1^*(\sigma)$ remains peaked at $z=1$, the distribution for $r=r_2^*(\sigma)<r_1^*(\sigma)$ widens and can even exhibit a maximum inside the support at large $\sigma$ (Figs. \ref{fig:rho_l_vs_z}b, \ref{fig:rho_l_vs_z}c and \ref{fig:rho_l_vs_z}d). In Fig. \ref{fig:rho_l_vs_z}d, the two optimal rates differ by more than one order of magnitude and produce radically different last resetting point distributions.
In summary, the two equivalent optimal strategies are very different. 

The truncated exponential distribution leads to qualitatively similar results. In this case the most probable position of $\rho_l(z)$ is given by
\begin{equation}
\hat{z}_l=
\begin{cases}
\mu\quad {\rm if}\ r <\frac{2D}{\sigma^2}\\
\\
1\quad {\rm if}\ r > \frac{2D}{\sigma^2}.
\end{cases}
\label{hatz2}
\end{equation}
Once again, a large drop of the optimal rate from $r^*_1$ to $r_2^*$ in Fig. \ref{fig:jumps_r_T} implies a significant broadening of the last resetting point distribution (not shown). 

\begin{figure*}
\centering
\begin{minipage}[b]{.45\textwidth}
\includegraphics[width=\linewidth]{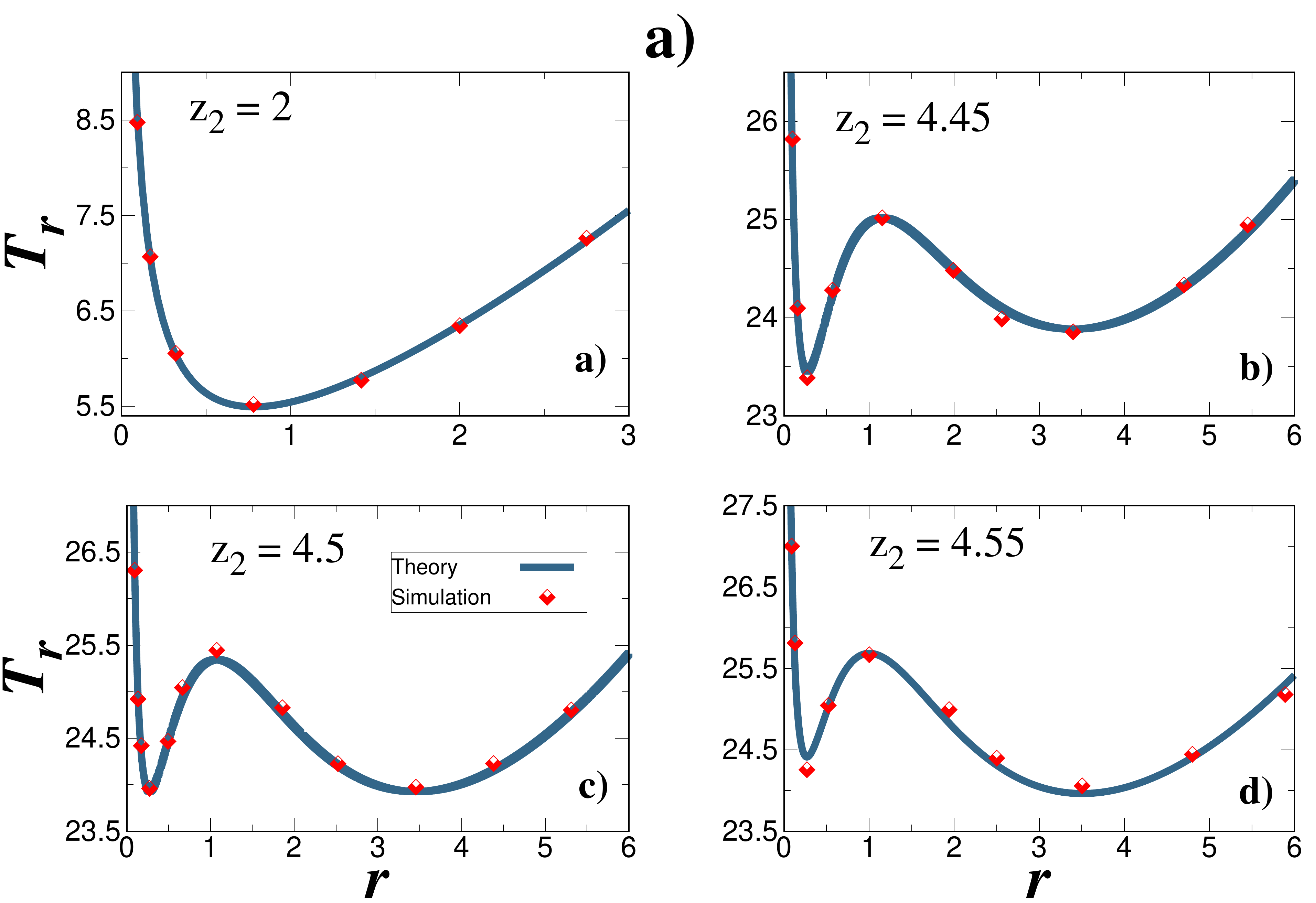}
    \label{fig:Tvsr}
\end{minipage}
\quad
\begin{minipage}[b]{.45\textwidth}
\includegraphics[width=\linewidth]{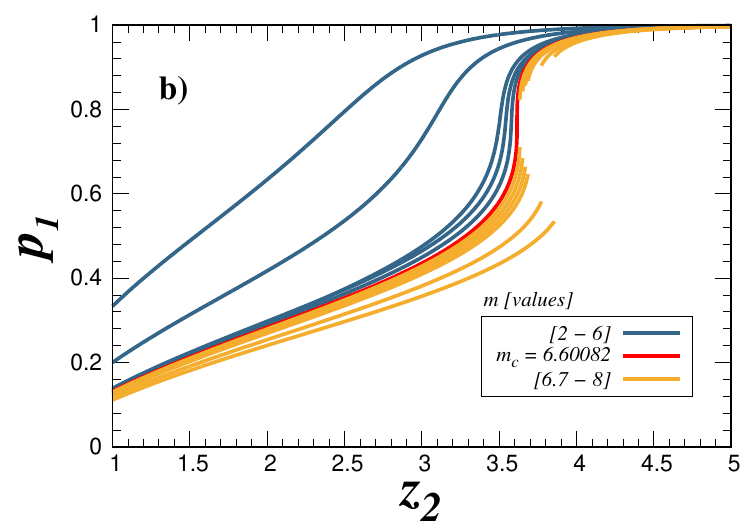} \label{fig:probvsxr}
\end{minipage}
\caption{(a) Transition in the global minimum of the AMFPT for the two-site resetting problem ($m=12.28$). The red markers are the result of numerical simulations. (b) Probability that the last diffusive path has started from the site closer to the target, given by Eq. \eqref{pdf-21}, for several values of $m$, and evaluated at the optimal rate $r^{\ast}$.} \label{fig:prob_vs_xr_Tvsr}
\end{figure*}

\subsection{Two-point resetting revisited}\label{sec:two_point_revisited}
We now address the last resetting point density in the two resetting points (TRP) problem, whose optimal properties were studied in \cite{two_resetting_points} and which we summarize below. Two resetting sites are located at the dimensionless positions $z_1=1$ (by convention) and $z_2>1$. 
The PDF of the resetting position, $\rho(z)$, is given by
\begin{equation}
\rho(z) = \frac{1}{1+m}\delta(z-1) + \frac{m}{1+m}\delta(z-z_2), \label{pdf-18}
\end{equation}
where $1/(1+m)$ and $m/(1+m)$ represent, respectively, the probabilities of choosing the positions $z_1$ and $z_2$ when a resetting event occur. The parameter $m\in(0,\infty)$ thus represents the relative weight of the further site at $z_2$. 

In \cite{two_resetting_points}, we showed that if $m>m_c=6.600817...$, the optimal resetting rate minimizing Eq. (\ref{pdf-4}) underwent a discontinuous transition at a certain value of $z_2$ denoted as $z_t(m)$. For illustration, in Fig. \ref{fig:prob_vs_xr_Tvsr}a we have set $m=12.28>m_c$ and displayed the variations of the AMFPT with $r$ for several increasing values of $z_2$. This figure is similar to Fig. \ref{fig:mfpt_gauss_expo} of the TGD and TED cases, where the parameter $\mu$ is analogous to $z_2$.
The red rhombuses of Fig. \ref{fig:prob_vs_xr_Tvsr}a are results from numerical simulations, which confirm our theoretical findings and the existence of two minima in the AMFPT.

However, an important difference can be noticed: as $z_2$ crosses $z_t(m)$ from below, the location of the global minimum jumps from $r_2^*$ (the lower rate) to $r_1^*$, and not from $r_1^*$ to $r_2^*$ as in the TGD and TED cases. Hence, the jump in $r^*$ is {\it upward}, and not downward as previously.

Following the same steps as in Section \ref{sec:GL}, the critical parameters $(r_c,z_c,m_c)$ are determined by solving the system of equations $\partial^n{\cal T}_r/\partial r^n|_{r_c,m_c,z_c}=0$ with $n=\{1,2,3\}$, which yields $r_c=1.4018...$, $m_c=6.6008...$ and $z_c=3.6158...$ \cite{two_resetting_points}. Near the critical point, we can use the same general Ginzburg-Landau expansion (\ref{gl-3}). To describe the shapes of Fig.  \ref{fig:prob_vs_xr_Tvsr}a, one must now choose the coefficients as
\begin{equation}
\label{a2trp}
a_2=-\alpha_2(m-m_c),\quad {\rm with}\ \alpha_2>0, 
\end{equation}
and 
\begin{equation}
\label{a1trp}
a_1=-\alpha_1(z_2-z_t(m)),\quad {\rm with}\ \alpha_1>0. 
\end{equation}
Notice the change of sign in Eq. (\ref{a1trp}) compared to Eq. (\ref{a1tgd}). Here, before the transition, when $z_2<z_t(m)$, $a_1$ is positive, thus the lower resetting rate is favored by the linear term in the minimization of Eq. (\ref{gl-3}). (In a magnetic analogy, this is similar to having a negative external field.) 

Fixing $m$ large enough, contrary to Fig. \ref{fig:jumps_r_T} for the TGD and some TED cases,  $r^*$ is non-monotonous (see Fig. 6 of \cite{two_resetting_points}): it  first decreases with $z_2$, then {\it increases} at large $z_2$ (with or without a discontinuity from $r_2^*$ to $r_1^*$) and tends to a finite value at $z_2=\infty$ \cite{two_resetting_points}. This behavior can be explained qualitatively as follows: for $z_2\sim 1$, the effect of the site $z_1$ is small and the problem is similar to having resetting to $z_2$ only, for which we know that $r^*$ decays as $1/ z_2^2$ \cite{evans_diffusion_2011}. If $z_2$ is too far, however, diffusive paths starting from $z_2$ take too long to reach the origin. It is thus convenient to take advantage of the closer site by choosing the time interval between consecutive resetting events ($1/r$) of the order of the diffusion time from $z_1$, which is short. Hence $r^*$ increases. These considerations suggest the following picture: as $z_2$ varies from $1$ to $\infty$, $r^*$ first belongs to the \lq\lq entropic" regime and then transits to the \lq\lq energetic" one. This scenario is opposite to the TGD and TED cases. The main difference resides in the fact that, here, the probability to reset close to the target (at $z=1$) remains constant and does not decay to $0$ when $z_2\to\infty$. 

The change of regime in the optimal strategy can be quantified by calculating the last resetting point distribution,
substituting Eq. \eqref{pdf-18} into Eq. \eqref{pdf-17}. The normalization constant is ${\cal C}=(m+1)/(e^{-\alpha_{0}} + m e^{-\alpha_{0}z_2})$ and $\rho_{l}(z)$ given by
\begin{eqnarray}
    \rho_{l}(z) &=& \frac{\delta(z-1) + m\delta(z-z_2)}{e^{-\alpha_{0}} +  me^{-\alpha_{0}z_2}}e^{-\alpha_{0}z}. \label{pdf-19}\\
    &\equiv& p_1\delta(z-1)+(1-p_1)\delta(z-z_2), 
\end{eqnarray}
where $p_1$ represents the probability that the last resetting point before absorption is the closer one.
By identification, 

\begin{equation}
    p_1(r,z_2,m)= \frac{1}{1 + me^{-\alpha_0( z_2-1)}}, \label{pdf-21}
\end{equation}
If there is only one resetting point ($m=0$), we recover the trivial result $p_1 = 1$. In the opposite limit, $\lim_{m\to\infty}p_{1}=0$, as the particle can only reach the target from $z_2$.

For several fixed values of $m$, Fig. \ref{fig:prob_vs_xr_Tvsr}b shows the last resetting point probability at optimality, i.e., $p_{1}(r^{\ast}(z_2,m),z_2,m)$, as a function of $z_2$. 
Clearly, this quantity increases monotonically with $z_2$. Fixing $m>1$, for $z_2 \sim 1$ one observes that $p_1<1/2$ meaning that the successful diffusive paths are more likely to emanate from the most probable site $z_2$. But as $z_2$ increases, the most effective paths increasingly come out from the closer site $z_1 = 1$ and $\lim_{z_2\to \infty}p_1(z_2)=1$. The blue curves ($m<m_c$) show a gradual cross-over between these two regimes, whereas in the yellow curves ($m>m_c$) the transition is abrupt and occurs at $z_2=z_t(m)$. 
When $m$ is very large, $p_1$ jumps from a small value $\sim 1/(1+m)$ to a value very close to unity. This case illustrates once again how different can be the two equally optimal strategies at $z_2=z_t(m)$.

\section{Conclusions}\label{sec:concl}
We have studied first passage observables of one-dimensional diffusion subject to resetting to random positions distributed on compact supports. Notably, the optimal resetting rate $r^{\ast}$ that minimizes the averaged mean first-passage time at a target located outside the support can exhibit a singularity and discontinuous transitions in parameter space. This phenomenology differs from standard single-site resetting and can be described generically by a Ginzburg-Landau approach, where the existence of the critical point (similar to a \lq\lq liquid-gas" point) obeys a simple general criterion. We have illustrated these results by considering truncated Gaussians and truncated exponential distributions of resetting points, adding to results previously obtained with two resetting points  \cite{two_resetting_points}.  

Distributed resetting exhibits rich behaviors when the most probable resetting positions $z$ are in the middle zone of the support, while the positions closer to the target (at the edge) are less likely. The optimal rate $r^{\ast}$ can exhibit a variety of features, depending on the shape of the distribution $\rho(z)$ of the resetting points. For example, $r^{\ast}$ decreases monotonously with the mean resetting distance in the Gaussian case (as in single-site resetting), while it can be non-monotonous for the exponential and the two-resetting site distributions.

A discontinuous or \lq\lq first order" transition occurs for $r^*$ as the first moment of $\rho(z)$ is varied (while the variance is fixed above its critical value). The discontinuity is due to the presence of two local minima in the AMFPT as a function of $r$, which become equivalent (or \lq\lq coexist") at the transition point. The corresponding rates ($r_2^*$ and $r_1^*>r_2^*$) represent slow and fast resetting strategies, and their values differ widely ($r_1^*\gg r_2^*)$ as the variance of $\rho(z)$ increases. 

Due to the discontinuity of the transition, notice that
one may anticipate hysteresis effects if certain experimental/numerical protocols are used in the determination of $r^*$. As exemplified by Fig. \ref{fig:mfpt_gauss_expo}a, assume that an operator has measured $r^*(\mu)=r_1(\mu)$ for a certain value of the parameter $\mu$, and locally looks for the next optimal value in the vicinity of $r^*(\mu)$ (e.g., by using steepest descent) as $\mu$ is slightly increased to $\mu+\Delta\mu$. 
The optimal rate obtained in this way, $r_1(\mu+\Delta\mu)$, may no longer be the global minimum (located at $r_2$, as in Fig. \ref{fig:mfpt_gauss_expo}c). The jump from $r_1$ to $r_2$ will be delayed and will occur only when 
the local minimum at $r_1$ disappears (or vice versa), causing an optimization depending on the path followed in parameter space, and therefore hysteresis.

We further interpreted our results by analyzing the probability density $\rho_{l}(z)$ of the last resetting position used by the particle before absorption. Upon a normalization constant, $\rho_{l}(z)$ is given by $\rho(z)$ weighted by an attenuation factor $e^{-\sqrt{r/D} |z|}$ that decays with the distance $|z|$ to the target. The distribution $\rho_l(z)$ identifies the resetting zone that is actually efficient for finding the target, and two qualitative regimes can be observed: 
the efficient zone can be the one nearest to the target (\lq\lq energetic" regime), or where the most probable resetting positions are (\lq\lq entropic" regime). The optimal rate $r^*$ can be controlled by either regime, depending on the distribution $\rho(z)$ chosen. As the mean resetting distance increases and crosses the transition value, the optimal rate usually jumps from the energetic to the entropic regime, but for the two-point distribution it is the other way around. The co-existing rates $r_1^*$ and $r_2^*$ are equally efficient for the target search, but they can be based on very different resetting zones of the support. 

In summary, search processes based on resetting to random positions that are distributed on a compact support can exhibit co-existence between two optimal resetting rates, corresponding to two markedly different strategies.  
When the mean and variance of the distribution are located on the co-existence curve, the two strategies are equally effective and are responsible for a discontinuity in the optimal rate. It would be interesting to extend these results to situations where there is a cost for resetting \cite{evans2018effects,sunil2023cost, sunil2024minimizing,olsen2024thermodynamic,gupta2025optimizing}, for instance by considering a refractory period after each reset.

\vspace{0.5cm}
\noindent {\bf{Data Availability}} The data that support the findings of this study are available
from the corresponding author upon reasonable request.

\vspace{0.5cm}
\noindent {\bf{Acknowledgments}} We acknowledge the financial support provided by SECIHTI, scholarship No. 4018217 and grant Ciencia de Frontera No. CBF-2025-I-3987. D.B. also acknowledges support from CONACYT (now SECIHTI) grant Ciencia de Frontera 2019/10872.

\bibliography{biblio}

\end{document}